\documentstyle[epsf,12pt]{article}
\title{Rescaling Relations between Two- and Three-Dimensional Local Porosity
Distributions for Natural and Artificial Porous Media}                
\author{B. Virgin$^{a}$, E. Haslund$^{a}$ and R. Hilfer$^{a,b,c}$}
\date{}
\begin{document}                
\maketitle                
{\small \noindent
 $^{a}$Department of Physics, University of Oslo, 0316 Oslo, Norway\\
 $^{b}$Institut f\"{u}r Physik, Universit\"{a}t Mainz, W-55099 Mainz, Germany\\
 $^{c}$ICA-1, Universit\"{a}t Stuttgart,
 Pfaffenwaldring 27, 70569 Stuttgart, Germany}
\begin{abstract}
Local porosity distributions for a three-dimensional porous medium and local 
porosity
distributions for a two-dimensional plane-section through the medium are generally 
different. However, for homogeneous and isotropic media having finite
correlation-lengths, 
a good degree of correspondence between the two sets of local porosity
distributions 
can be obtained by rescaling lengths, and the mapping associating corresponding
distributions can be found from two-dimensional observations alone.
The agreement between associated distributions is good as long as the 
linear extent of the measurement cells involved is somewhat larger than the 
correlation length, and it improves as the linear extent increases. 
A simple application of the 
central limit theorem shows that there must be a correspondence in the limit of 
very large measurement cells, because the distributions from both sets 
approach normal distributions. A normal distribution has two independent
parameters: the mean and the variance. If the sample is large enough,
LPDs from both sets will have the same mean. Therefore corresponding distributions
are found by matching variances of two- and three-dimensional local porosity
distributions.
The variance can be independently determined from correlation functions. Equating
variances 
leads to a scaling relation for lengths in this limit. Three particular systems
are examined in order to show that this scaling behavior
persists at smaller length-scales.
\end{abstract}
\newpage

\section{Introduction}                
The study of porous and heterogeneous media is scattered throughout
many fields of science, including mathematics, solid state physics and
chemistry, materials science, geology, hydrology, environmental technology,
petroleum engineering and separation technology. The central problem is
the specification of the random microstructure, which is needed to predict
macroscopic physical properties ( see 
\cite{Landauer}\cite{Mochan}\cite{Hilfer4}\cite{Sahimi}\cite{Lafait}
for overviews).

A complete specification of the random microstructure is both impractical
and unnecessary. It is therefore important to have a general statistical
description of the microstructure available. Such a description should meet
four criteria: it should be well-defined in terms of geometrical quantities,
it should involve only experimentally accessible parameters, it should be
of economical size and it should be usable in exact or approximate solutions
of the underlying equations of motion. Currently there are only two statistical
methodologies available which fulfill all four requirements; these are 
correlation functions
\cite{Debye1}\cite{Debye2}\cite{Weissberg}\cite{Torquato1}\cite{Torquato2}\cite{Rikvold1}\cite{Rikvold2}\cite{Stell}\cite{Berryman1}\cite{Berryman2}
and local geometry distributions
\cite{Hilfer1}\cite{Hilfer2}\cite{Boger}\cite{Hilfer3}\cite{Hilfer4}.

The recently developed
local geometry distributions are a functional generalization of the
correlation-function approach\cite{Hilfer4}. The aim of the present paper 
is to increase the practical applicability of local geometry distributions.
Interest in these distributions arises from their potential 
use in distinguishing between different microstructures\cite{Boger}.  
Increasing their practical applicability is particularly important because
local geometry distributions can be
used in mean field theory calculations of the
dielectric and transport properties of various porous media, for instance 
porous rock filled with brine\cite{Hilfer1}\cite{Hilfer2}. 
Preliminary comparisons of the theory with experimental results have 
already been
carried out, and show good agreement\cite{Hilfer3}\cite{Haslund}.

Two main objectives will be pursued in the present paper. First,
we show the feasibility of obtaining three-dimensional local porosity
distributions(LPDs), by 
presenting data from a real sandstone specimen. Second, we
show to what extent three-dimensional LPDs can be obtained from purely
two-dimensional observations. 
The dependence of LPDs upon the dimensionality of the system is a 
consequence of their dependence on 
a measurement cell or observation region over which geometric
observables, such as porosity, are averaged. The possibility of extracting 
three-dimensional distributions from two-dimensional data arises from limit
theorems which guarantee that the asymptotic LPDs become independent of the
shape of the measurement cell\cite{Hilfer4}.

We begin our discussion with a definition of LPDs. Next we present
a simple argument based on the central limit theorem showing that for
large measurement cells, three-dimensional LPDs can be mapped onto
two-dimen\-sion\-al
LPDs by a simple rescaling of lengths. Then, we show that this mapping
becomes exact for the percolation model. Subsequently, we analyse the
overlapping sphere (or continuum percolation) model, where the mapping
holds  only asymptotically. Finally we present the first measurement of
three-dimensional LPDs. These are obtained from the pore-space 
reconstruction of a real sandstone specimen.

The main result of our investigation is the fact that the mapping 
between two- and three-dimensional LPDs can be determined from purely
two-dimen\-sion\-al observations if the medium is homogeneous and isotropic.
For general anisotropic media however, the family of three-dimensional
LPDs can only be determined from a three-dimensional measurement.

\section{Basic Concepts and an Expression for Variance}

The discretized porous medium is considered to occupy a set
${\mathbf{S}} \subset {\mathbf{Z}}^{d}$, where ${\mathbf{Z}}$ is the set
of integers. Pore-space is denoted by $\mathbf{P}$, 
where $\mathbf{P \subset S}$. Considering for simplicity only two-component media,
we define matrix-space $\mathbf{M}$ as the complement of $\mathbf{P \mbox{ in } S}$,
$\mathbf{M = S - P}$. The characteristic function of a set $\mathbf{A}$
is
\[
\mathbf{
\chi_{\mathbf{A}}(r) = \left\{ \begin{array}{ll}                
				1, & {\mathbf{r \in A}} \\         
				0, & \mbox{otherwise}                
			       \end{array}                
		       \right.                 
}
\]        
where ${\mathbf{r \in Z}}^{d}$. To describe the state of each point
in the system, we label each point in ${\mathbf{S}}$ with an index $i$ and
define the random variables
\[
  X_{i} \equiv X(\mathbf{r}_{i}) = \chi_{\mathbf{P}}(\mathbf{r}_{i})
\]
A specific geometry is then specified by the $N$-tuple
\[
  g = (X_{1}, \ldots ,X_{N}),\mbox{\ \ \ \ } N = |{\mathbf{S}}| 
\]
where $|{\mathbf{S}}|$ is the number of elements in $\mathbf{S}$.

The probability space for the system is the triple $({\mathbf{G}},{\cal{E}},
P)$. $\mathbf{G}$ is the sample space consisting of all possible geometries $g$. 
The event space $\cal{E}$ is a 
Boolean algebra of subsets of $\mathbf{G}$. The probability $P$ of finding a
geometry\footnote{Strictly speaking, $P$ is defined on 
sets of geometries
${\mathbf{E}} \in {\cal E}$. However, we assume that $\{g\} \in {\cal E}$
for all $g \in {\mathbf{G}}$.} $g$ is a finitely
additive non-negative set function defined on $\cal{E}$, such that 
$P(\emptyset) = 0$ and $P({\mathbf{G}})=1$. A stochastic
medium is homogenous if $P$ is invariant under translations.
Isotropy is defined analogously as invariance under rotations. 
The expectation value of a random variable $f(g)$ is
\[
\langle f(g) \rangle = \sum_{g \in \mathbf{G}} f(g)P(g)
\]
and in the following, we will use the notation
\begin{eqnarray*}
  S_{n}({\mathbf{r}}_{1},\ldots, {\mathbf{r}}_{n}) & = & 
  \langle \prod_{i=1}^{n} 
  \chi_{\mathbf{P}}({\mathbf{r}}_{i})\rangle \\
  C_{n}({\mathbf{r}}_{1},\ldots, {\mathbf{r}}_{n}) & = & 
  \langle \prod_{i=1}^{n}
  \chi_{\mathbf{P}}({\mathbf{r}}_{i}) - \langle 
  \chi_{\mathbf{P}}({\mathbf{r}}_{i}) \rangle \rangle
\end{eqnarray*}
for the moment functions $S_{n}({\mathbf{r}}_{1},\ldots, {\mathbf{r}}_{n})$
and cumulant functions $C_{n}({\mathbf{r}}_{1},\ldots, {\mathbf{r}}_{n})$
of order $n$. For homogeneous media, 
$S_{1}({\mathbf{r}})=\langle \phi \rangle$ is the expected bulk porosity.

Given a measurement cell denoted as $\mathbf{K}\ ({\mathbf{K}} \subset
{\mathbf{S}})$, we define the local 
porosity of $\mathbf{K}$ as the random variable
\[
  \phi({\mathbf{K}}) = {\mathbf{\frac{|P\cap K|}{|K|}}} = \frac{1}{|{\mathbf{K}}|}
  \sum_{{\mathbf{r \in Z}}^{d}} \chi_{\mathbf{P}}(\mathbf{r})
  \chi_{\mathbf{K}}(\mathbf{r})
\]
The local porosity distribution(LPD)\cite{Hilfer1} for $\mathbf{K}$
can be defined as\cite{Hilfer4}
\[
  \mu(\phi;\mathbf{K}) = \langle \delta(\phi - \phi(\mathbf{K}))\rangle
\]
where $\delta(\phi - \phi(\mathbf{K}))$ is the Dirac delta function,
and the first moment of this LPD is
\[
  \overline{\phi\left({\mathbf{K}}\right)}=\int_{0}^{1} \phi\,
  \mu(\phi;{\mathbf{K}})\, d\phi
\] 
Using the definition of $\mu$, and changing the order of summation gives
\[
  \overline{\phi\left({\mathbf{K}}\right)}
  =  \sum_{g \in \mathbf{G}} \left( \int_{0}^{1}\phi\,
  \delta(\phi-\phi({\mathbf{K}}))\, d\phi
  \right) P(g) =  \sum_{g} \phi({\mathbf{K}})P(g) =  
  \langle \phi(\mathbf{K}) \rangle
\]
\[
  =  \left\langle\frac{1}{|\mathbf{K}|}\sum_{\mathbf{r \in Z}^{d}} 
    \chi_{\mathbf{P}}({\mathbf{r}})\chi_{\mathbf{K}}({\mathbf{r}})\right\rangle
   = \frac{1}{|\mathbf{K}|}\sum_{\mathbf{r}}\langle \chi_{\mathbf{P}}(\mathbf{r})
    \rangle \chi_{\mathbf{K}}(\mathbf{r})
\]
where we have also used the definition of $\phi(\mathbf{K})$.
Thus, we obtain the expression 
\begin{equation}
  \overline{\phi\left({\mathbf{K}}\right)} =
  \int_{0}^{1}\phi\,\mu(\phi;{\mathbf{K}})\,d\phi = 
  \langle \phi({\mathbf{K}}) \rangle =
  \frac{1}{|\mathbf{K}|}
  \sum_{\mathbf{r}} S_{1}(\mathbf{r})\chi_{\mathbf{K}}(\mathbf{r})
\label{mean}
\end{equation}
For homogeneous media, $S_{1}=\langle \phi \rangle$, the bulk porosity, 
so that $\overline{\phi\left({\mathbf{K}}\right)} = \langle \phi \rangle$.
Similarly, we can show that the local porosity variance 
$var[\phi\left({\mathbf{K}}\right)] =\mbox{\-}
\int_{0}^{1}\mbox{\-}\,\left(\phi-\overline{\phi\left({\mathbf{K}}\right)}\right)^{2}
\mbox{\-}\mu(\phi;{\mathbf{K}})\,d\phi$ is given by
\begin{equation}
  var\left[\phi(\mathbf{K})\right] = \frac{1}{|\mathbf{K}|^{2}}
  \sum_{{\mathbf{r,r^{'} \in Z}}^{d}} C_{2}({\mathbf{r, r}}^{'})
  \chi_{\mathbf{K}}({\mathbf{r}}) \chi_{\mathbf{K}}({\mathbf{r}}^{'})
\label{var}
\end{equation} 
If the medium is homogeneous we can perform one of the sums. In this case
$C_{2}({\mathbf{r,r}}^{'}) = C_{2}({\mathbf{0,r}}^{'}-{\mathbf{r}})$.
Defining
$\mathbf{y = r}^{'}-\mathbf{r}$, and using the fact that $\chi_{\mathbf{K}}
(\mathbf{r+y}) = \chi_{\mathbf{K-y}}(\mathbf{r})$, where $\mathbf{K-y}$ is the
set $\mathbf{K}$ translated by the vector $-\mathbf{y}$, we obtain
\begin{equation}
  var\left[ \phi(\mathbf{K})\right] =  \sum_{{\mathbf{y \in Z}}^{d}} 
  \frac{|\mathbf{K \cap (K-y)}|}{|\mathbf{K}|^{2}} C_{2}({\mathbf{0,y}})
\label{homo.var}
\end{equation}
Note that the variance $var[\phi\left({\mathbf{K}}\right)]$ is simply
related to the granularity concept which is used for photographic 
materials\cite{O'Neill}. The relations above are applicable to infinitely
large realizations of homogeneous stochastic media.

\section{The Central Limit Theorem and Scaling}
Consider again the class of homogeneous media in which spatial correlations of the 
random geometry decay with a finite correlation-length $\xi$.
For such systems, we can apply the central limit 
theorem to measurements of local porsity for very large measurement cells. 
To do this, we partition a $d$-dimensional 
hypercubic measurement cell ${\mathbf{K}}$ of side $L$ into $n=(L/a)^{d}$ 
hypercubic subcells $\{{\mathbf{K}}_{i}\},\,i=1,2,\ldots,n$, of side $a$
so that
\[
  {\mathbf{K}} = \bigcup_{i=1}^{n} {\mathbf{K}}_{i}, \mbox{\ \ \ \ }
  {\mathbf{K}}_{i} \cap 
  {\mathbf{K}}_{j} = \emptyset \mbox{\ \ for \ } i \neq j
\]
and rewrite $\phi({\mathbf{K}})$ as the average
\[
  \phi({\mathbf{K}}) = \frac{1}{n}\sum_{i} \phi({\mathbf{K}}_{i}) 
\]
over all the subcell porosities $\phi({\mathbf{K}}_{i})$.
If we choose $L$ and $a$ such that $\xi \ll a \ll L$, then
$n \gg 1$, and the central limit theorem becomes applicable, because the 
$\phi({\mathbf{K}}_{i})$ are weakly correlated random variables.
The central limit theorem ensures the existence of a number
$\sigma$ such that
\[ \mu(\phi;{\mathbf{K}}) \simeq \frac{\left(L/a\right)^{d/2}}
                             {\sqrt{2\pi}\sigma}
    \exp \left[ -\frac{(\phi- \overline{\phi} )^{2}
      \left(L/a\right)^{d}}{2\sigma^{2}}\right]
\]
were the accuracy of the approximate equality improves as $L/a$ increases.
In the limit $L/a \rightarrow \infty$ subject to $a \gg \xi$, one has 
$\mu \rightarrow \delta(\phi - \overline{\phi})$, independent of dimension
and independent of ${\mathbf{K}}$. Hence, for $L/a$ sufficiently large
but still finite, the form of $\mu(\phi;{\mathbf{K}})$
depends only on the bulk porosity $\overline{\phi}$ and the variance
$var[\phi({\mathbf{K}})] = \sigma^{2}/(L/a)^{d}$, but not on the 
shape of $\mathbf{K}$.
If the medium is also isotropic, then
a sufficient condition for the equality of two such limiting distributions 
for $d$ and $d^{'}$-dimensional measurement cells in the same medium is thus
\begin{equation}                
\left(\frac{L^{'}}{a^{'}}\right)=                
\left(\frac{\sigma^{'}}{\sigma} 
\right)^{2/d^{'}}\,\left(\frac{L}{a}\right)^{d/d^{'}}             
\label{Approx.scaling}
\end{equation}                
because $\overline{\phi}$ will be the same in both cases.

Thus we find that for homogeneous isotropic systems with finite
correlation-length, two- and three-dimensional LPDs for large $L$ 
are simply related by a rescaling of lengths.
We now examine the site percolation model, for which the
scaling relation is exact.

\section{The Percolation Model}             
In the site percolation model, each
lattice site has a probability $\langle \phi \rangle$ of being occupied and 
$(1 - \langle \phi \rangle)$ of being empty that is independent of the other sites. 
Therefore, if we consider occupied sites to be pore and empty sites as matrix, 
$\mu(\phi; {\mathbf{K}})$ depends only on $\langle \phi \rangle$ and $|\mathbf{K}|$,
and is given by the binomial distribution. If the underlying lattice is hypercubic, 
then the LPD for a $d$-dimensional hypercubic measurement cell $\mathbf{K}$ 
of side $L$ is
\[
\mu(\phi_{i};{\mathbf{K}}) =                
\frac{L^{d}!}{\left[L^{d}\,\phi_      
{i}\right]!\:\left[L^{d}\,                
\left(1-\phi_{i}\right)\right]!}\;                
\langle\phi \rangle^{\,L^{d}\phi_{i}}\,(1-                
\langle \phi \rangle )^{\,L^{d}(1-\phi_i)} 
\] 
where $\phi_{i} = i/L^{d} \mbox{\ and\ } i= 0,1,\ldots,L^{d}$.                
Measurement cells consisting of the same number of points must have 
identical LPDs. Therefore, if ${\mathbf{K}}^{'}$ is a $d^{'}$-dimensional 
measurement cell of side $L^{'}$
\[
  \mu(\phi_{i};{\mathbf{K}}^{'}) = \mu(\phi_{i};{\mathbf{K}})
  \mbox{\ \ if\ \ } L^{'d^{'}} = L^{d}
\]

The scaling relation
\begin{equation}
   L^{'} = L^{d/d^{'}}
\label{Perc.scale}
\end{equation}                
is not solvable in ${\mathbf{Z}}$ for all $L,\,L^{'}$. Nevertheless,
there exist infinitely many solutions, since each positive integer $i$ 
yields the solution $L^{'} = i^{d} \mbox{ and } L = i^{d^{'}}$
Therefore, for the site percolation model, two- and three-dimendional LPDs 
are, within the constraints imposed by the discrete nature of the model, 
simply related by a rescaling of lengths.

\section{The Overlapping Sphere Model}
\begin{figure}
\epsfysize=325pt.
\epsffile{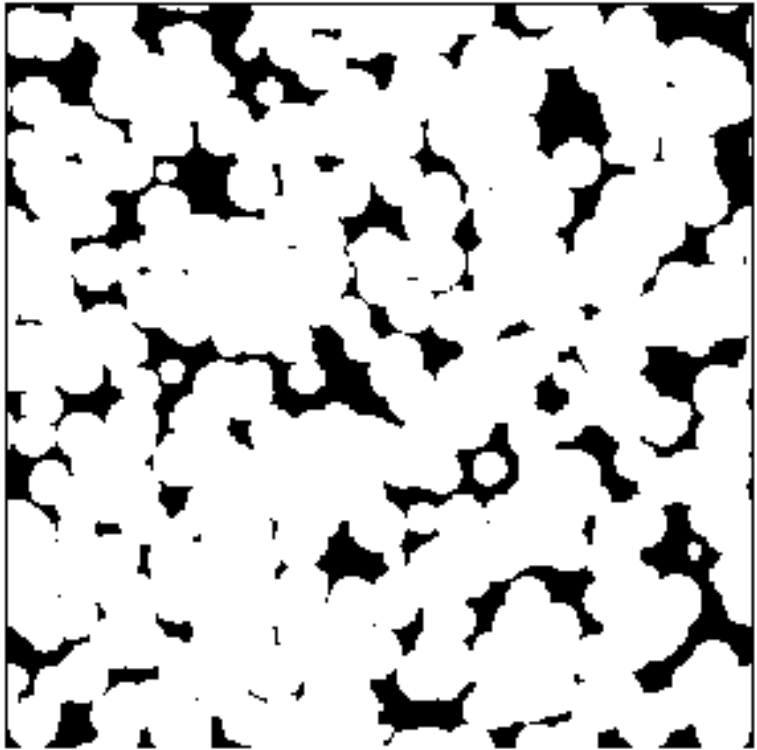}
\caption{A digitized plane-section from a 256x256x256-voxel digitized 
configuration of overlapping spheres, consisting of 8000 randomly distributed 
spheres, all of radius $r_{0} = 9.45$ voxels.}
\label{fig:pp0pgm}.
\end{figure}

An often useful grain model for porous media is obtained by randomly
distributing
overlapping spheres. The spheres represent grains of matrix material. A model
configuration is typically
created by randomly distributing point-centers with constant point-density in 
a given volume and attaching a sphere of radius $r_{0}$ to each point-center.
The system analyzed below was obtained
by randomly distributing 8000 point-centers in a volume of 
256x256x256 voxels. A value of $r_{0} = 9.45$ voxels has been used, and
periodic boundary conditions have been implemented.
The bulk porosity of the overlapping sphere model is $\langle \phi \rangle =
\exp \left[-\rho\frac{4}{3}\pi r_{0}^{3}\right]$, where $\rho$ is the
point-density. The point density and sphere
radius were chosen such that $\langle \phi \rangle = 0.1855$, matching
the porosity of a real sandstone specimen discussed below. A plane-section 
showing the degree of discretization is displayed in Figure~\ref{fig:pp0pgm}.

\begin{figure}
\epsfysize=460pt.
\epsffile{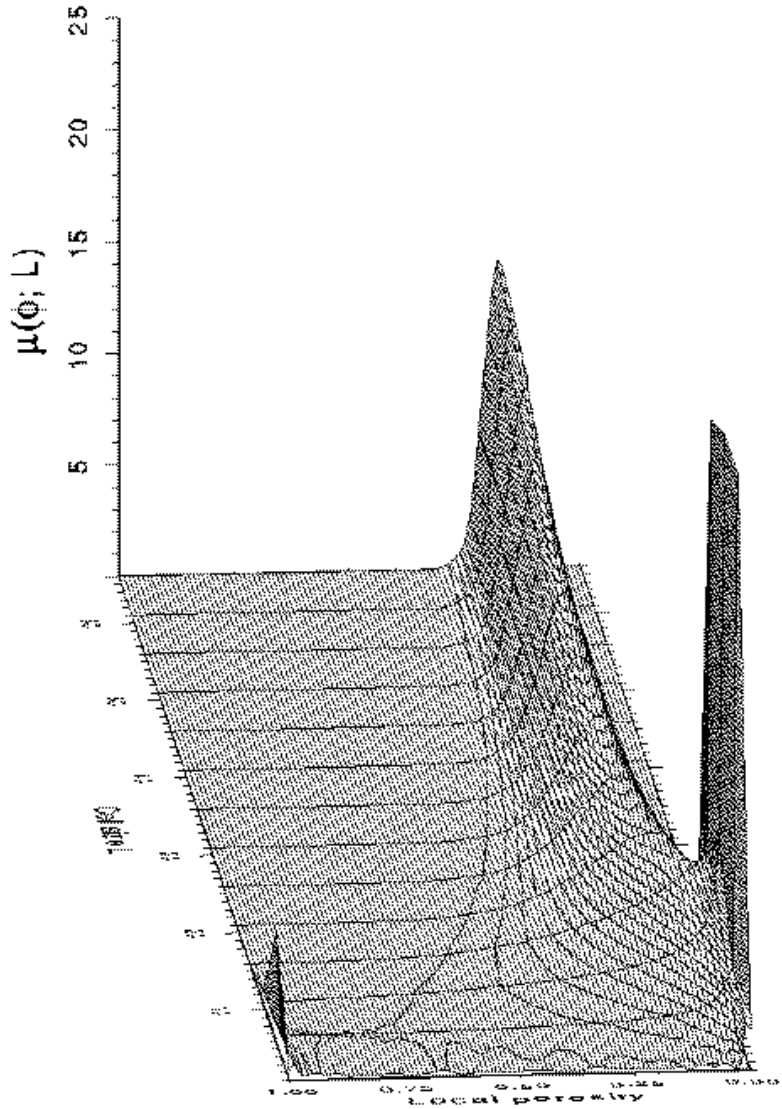}
\caption{The family of three-dimensional LPDs $\mu_{3d}(\phi;L)$ for 
$L \in \{1,2, \ldots,64\}$ for the overlapping sphere model. Note that the 
position of the saddle-point is near $L=18.9$, which is the correlation 
length for the system.}
\label{fig:p183dmu}
\end{figure}
\begin{figure}
\epsfysize=460pt.
\epsffile{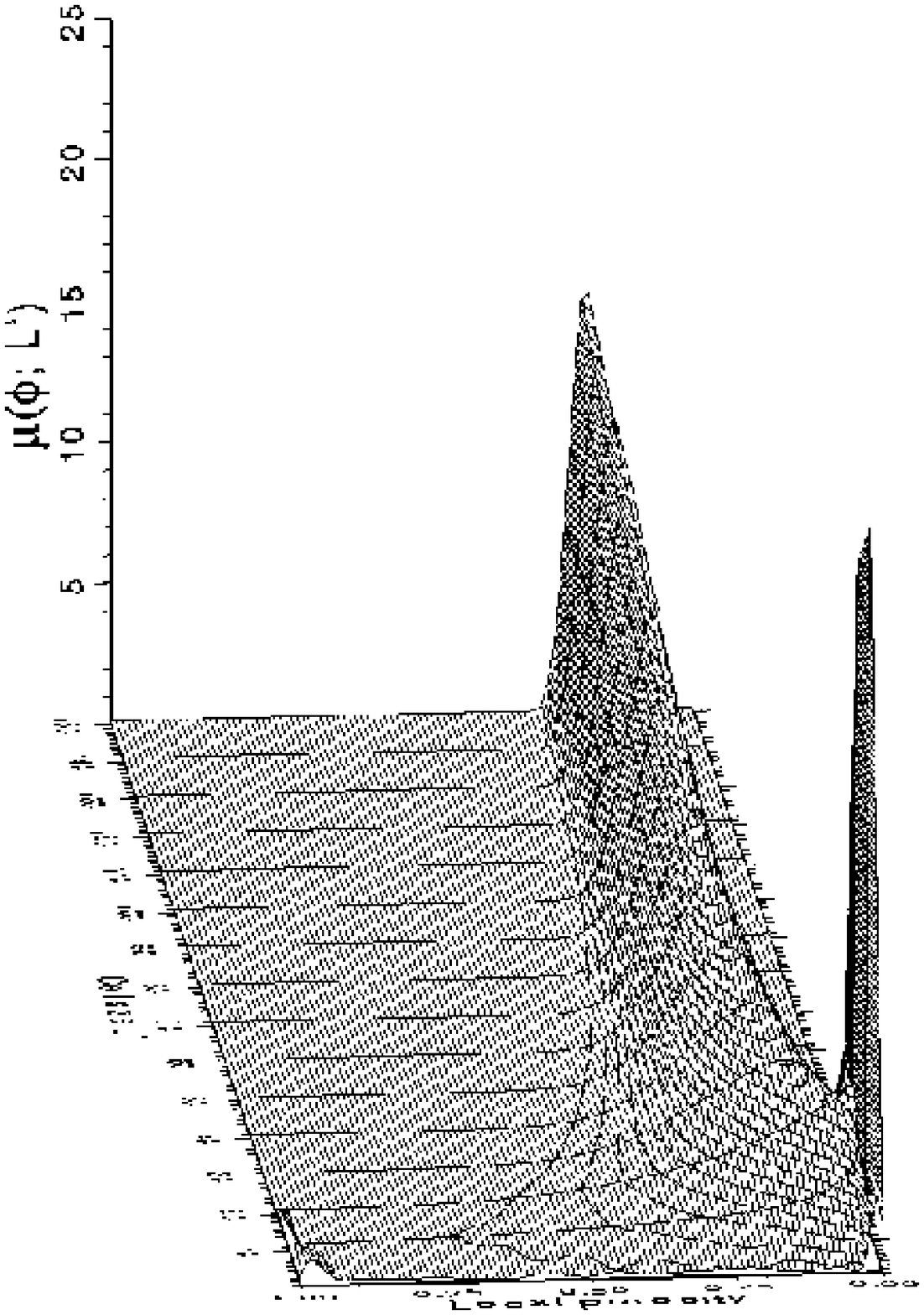}
\caption{The family of two-dimensional LPDs $\mu_{2d}(\phi; L^{'})$ for 
$L^{'} \in \{1,2,\ldots,150\}$. Although this surface is not the same as the
in the previous one, the previous surface can
be approximately recreated from this one by rescaling the $L^{'}$ axis.}
\label{fig:p182dmu}
\end{figure}
\begin{figure}
\epsfysize=450pt.
\epsffile{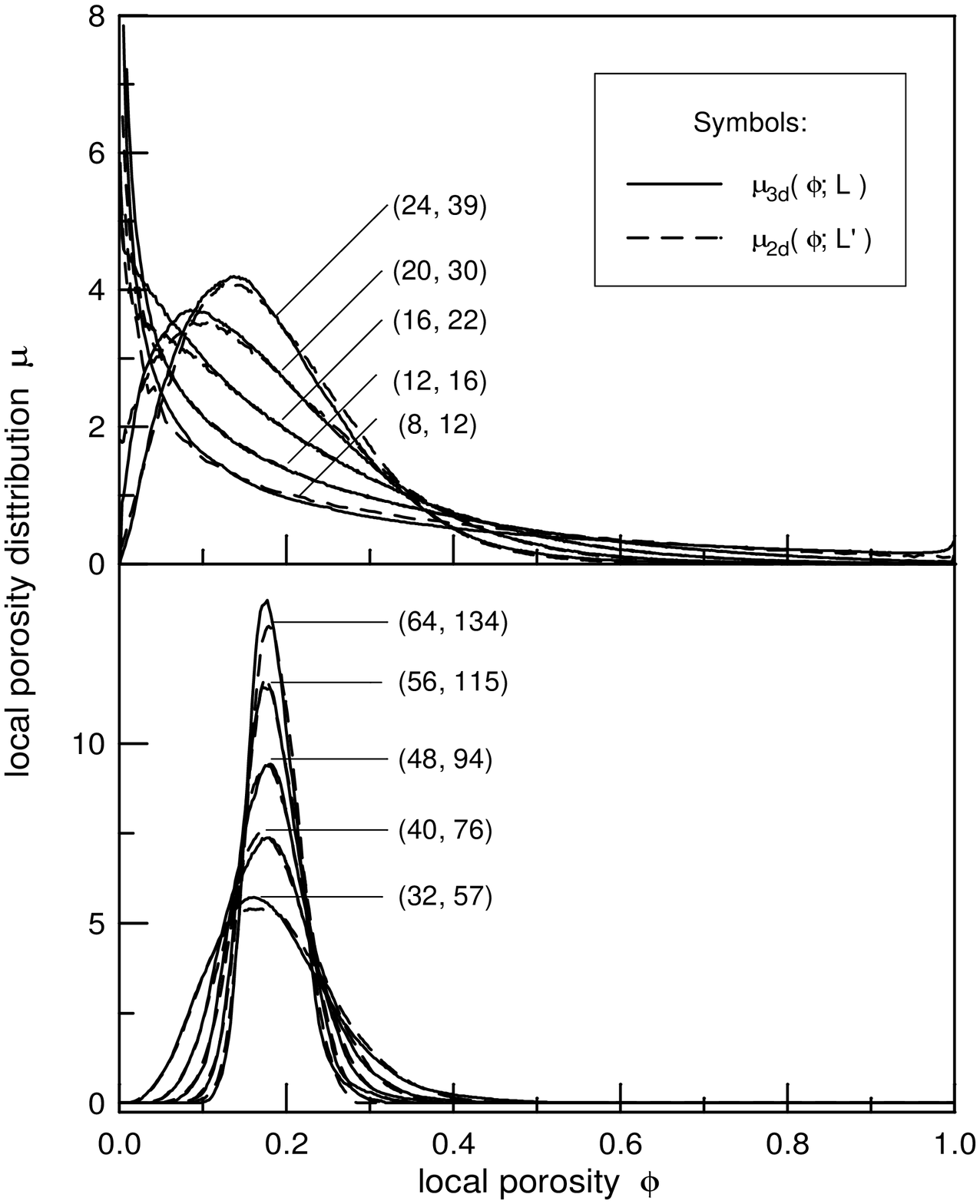}
\caption{A selection of matching two- and three-dimensional LPDs from 
the mapping obtained by minimizing $\Delta(L, L^{'})$. Solid lines indicate 
three-dimensional LPDs and dashed lines show two-dimensional LPDs. The 
side-length of the measurement cells for matched LPDs are shown on 
the figure as $(L, L^{'})$-pairs.}
\label{fig:p18corr}
\end{figure}
\begin{figure}
\epsfysize=245pt.
\epsffile{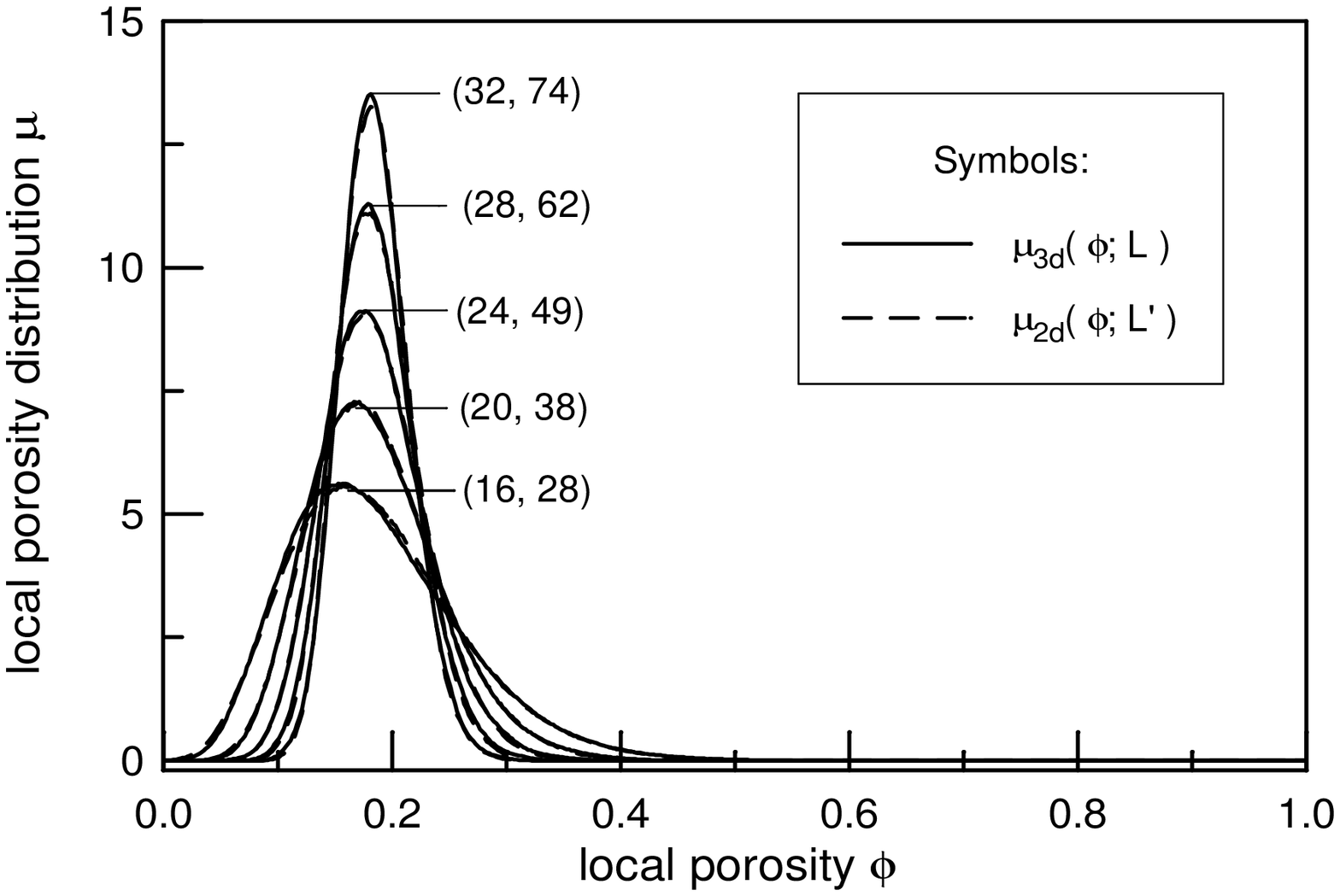}
\caption{Selected $\Delta(L,L^{'})$-matched distributions, similar to those
shown in 
the lower half of the previous figure, from an ensemble of 64000 spheres of
$r_{0} = 4.73$ voxels, within a volume of 256x256x256 voxels. The matching
$(L,L^{'})$-pairs are indicated in the figure.}
\label{fig:c2corr}
\end{figure}

Three-dimensional LPDs for $L \in \{1,2, \ldots, 64\}$ have been measured, 
and the resulting family of distributions is shown in Figures~\ref{fig:p183dmu}.  
Two-dimensional LPDs were obtained for $L^{'} \in \{1,2, \ldots, 150\}$ by
averaging the measured distributions for all cross-sections perpendicular to 
the z-axis. The results are displayed in Figure~\ref{fig:p182dmu}. To investigate
the degree 
of correspondence between two- and three-dimensional LPDs we calculate the
quantity 
\[
\Delta(L, L^{'}) = \int_{0}^{1} |\,\mu_{3d}(\phi;L)-\mu_{2d}(\phi;L^{'})\,|\,d\phi
\] 
as a quantitative measure of the deviation between two distributions.
For fixed $L$, we minimize $\Delta(L,L^{'})$ by varying $L^{'}$. The matched 
distributions selected by this criterion are shown in Figure~\ref{fig:p18corr}
for a choice 
of $(L, L^{'})$-values indicated in the figure for each matching pair.
The agreement between the matched two- and three-dimensional distributions is 
satisfactory. The agreement for large $L$ stipulated by the central limit
theorem improves with better system-statistics. This is seen from the matched 
LPDs shown in Figure~\ref{fig:c2corr}, for a system of 64000 instead of 8000 spheres,
having the same voxel-dimensions and bulk porosity. The poorer 
sphere-resolution of $r_{0} = 4.73$ pixels makes this configuration less
suitable for analyzing $\mu$ at smaller length-scales, 
so we shall not discuss this configuration further.

\begin{figure}
\epsfysize=350pt.
\epsffile{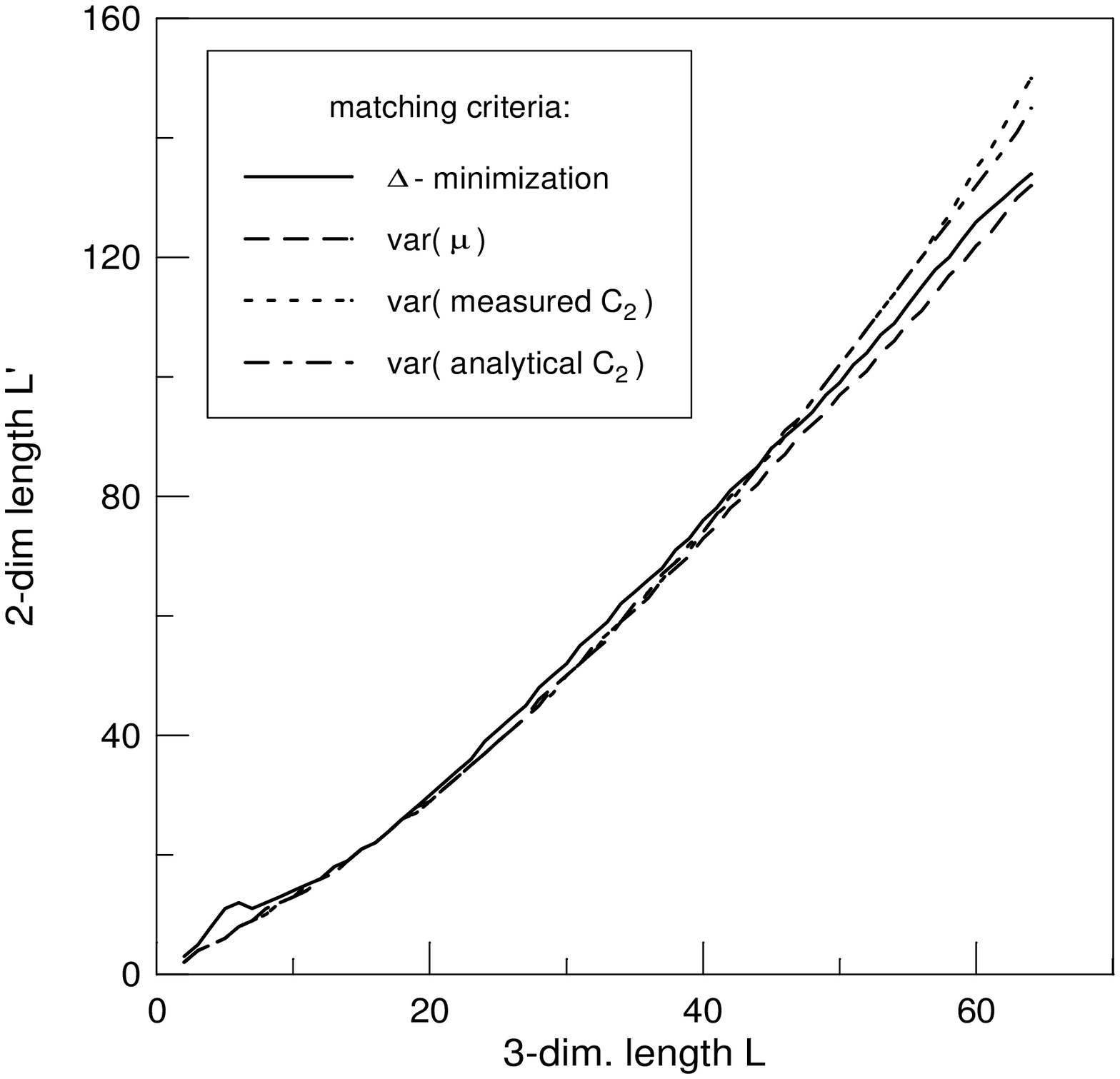}
\caption{The scaling relation $L^{'}(L)$ associating two- and three-dimensional
LPDs, as determined by  four different criteria. The solid line is based on
minimizing $\Delta(L,L^{'})$. The remaining curves are based on equating
variances: the dashed curve is obtained from the actual distribution variances,
the dotted curve and the dash-dotted curves are obtained from variances computed
from (\protect\ref{homo.var}) using measured and
analytical values for $C_{2}$
respectively.}
\label{fig:p18scale}
\end{figure}

The matching of distributions based on $\Delta(L,L^{'})$ is shown as the
solid line in Figure~\ref{fig:p18scale}. 
Another matching of distributions is obtained from the criterion of equal
variances established in section~3. The mapping $L^{'}(L)$ obtained from the
equal-variance criterion is shown as the dashed curve in 
Figure~\ref{fig:p18scale}. For lengths that are not too large relative to
the size of 
the system, both criteria give more or less the same rescaling of lengths
$L^{'}(L)$. Note, however, that the discretization can affect $L^{'}(L)$ at
small $L$.

\begin{figure}
\epsfysize=350pt.
\epsffile{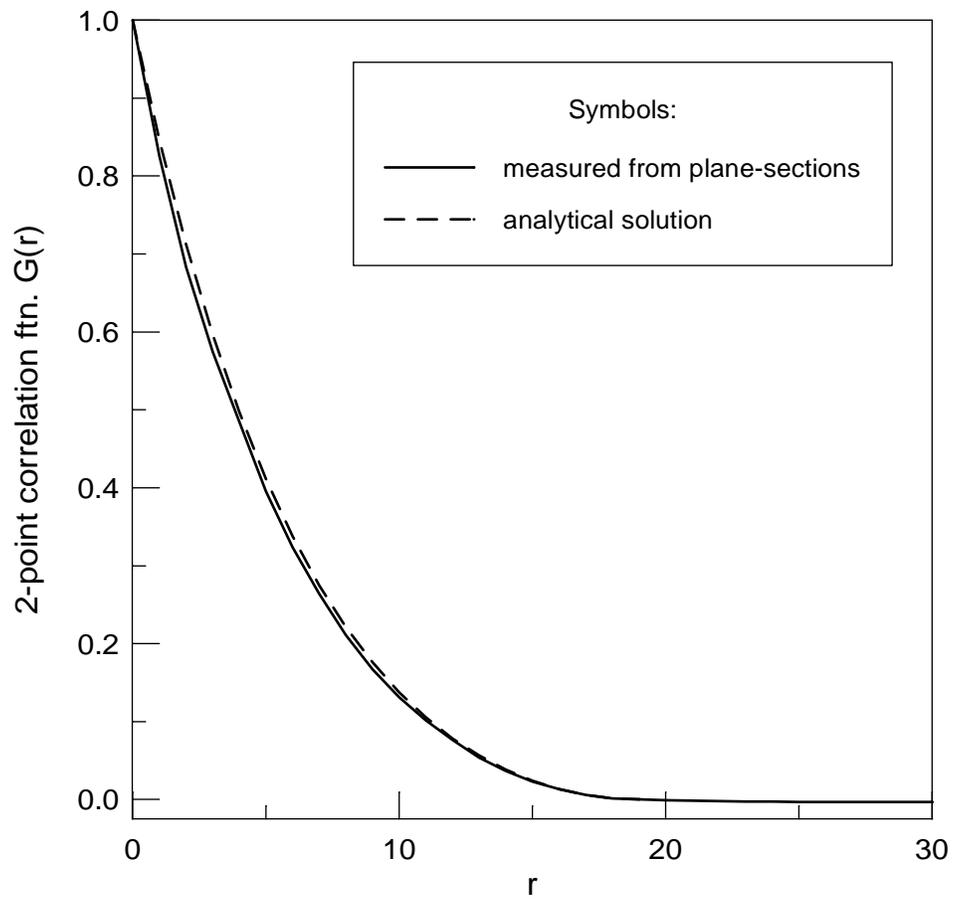}
\caption{The 2-point correlation function $G$ as a function of distance $r$. 
The solid curve shows measured values, while the dashed curve is the
analytical result for an infinite sample.}
\label{fig:p18g}
\end{figure}

In practical applications, it is important to estimate the three-dimensional
LPDs from observations on two-dimensional cross-sections alone. In the
matching above, however, it was assumed that the three-dimensional LPDs
are known. A way out of this dilemma is provided by equations (\ref{homo.var})
and (\ref{Approx.scaling}). The agreement between the dashed and the solid
lines in Figure~\ref{fig:p18scale} validates the matching criterion based
solely on the
variances expressed in equation (\ref{Approx.scaling}). On the other hand,
equation (\ref{homo.var}) allows us to estimate the variance from
$C_{2}({\mathbf{0,r}})$ measured on two-dimensional cross-sections as long
as the system is isotropic. The results of such a matching procedure are
shown as the dotted and the dash-dotted curves in Figure~\ref{fig:p18scale}.
The dotted
curve uses a direct measurement of $C_{2}$ from two-dimensional sections,
while the dash-dotted curve uses the exactly known analytical form of $C_{2}$.
The direct measurement of $C_{2}$ uses the same series of planar-sections
as the measurement of the two-dimensional LPDs; the correlation function is
first calculated in each plane and subsequently averaged over all planes.
The necessity to average over all planes, and the influence of discretization
effects were recently discussed systematically\cite{Coker}. The resulting
averaged two-point correlation function
\[
G({\mathbf{0,r}}) = \frac{C_{2}({\mathbf{0,r}})}{C_{2}({\mathbf{0,0}})}
= \frac{S_{2}({\mathbf{0,r}})-S_{1}({\mathbf{0}})S_{1}({\mathbf{r}})}
{S_{2}({\mathbf{0,0}})-S_{1}({\mathbf{0}})S_{1}({\mathbf{0}})}
\]
is shown in Figure~\ref{fig:p18g}. Because the model of overlapping spheres is
homogeneous and isotropic $G({\mathbf{0,r}})=G(r)$. The exact
analytical expression for $S_{2}({\mathbf{0,r}})$ is given as
\[
  S_{2}({\mathbf{0,r}}) = S_{2}(|{\mathbf{r}}|) =
  \exp \left[ -\rho\left(\frac{\pi}{3}d^{3} - V_{3}({\mathbf{B(0}},\frac{d}{2})
  \cap {\mathbf{B(r}},\frac{d}{2})\right) \right]
\]
$\rho$ is the density of point-centers, $d$ is the sphere-diameter,
$V_{3}({\mathbf{A}})$ is the volume of the set ${\mathbf{A}}$ and
\(
  {\mathbf{B(r}},c) = \left\{ {\mathbf{x : |x - r| \leq}} c \right\} 
\) is the ball of radius $c$ around $\mathbf{r}$. The exact $G(r)$ vanishes
for $r \ge d$, and is shown as the dashed curve in Figure~\ref{fig:p18g}. Visible 
discrepencies between the measured and analytical curves are due to
insufficient system-statistics.

The satisfactory agreement of the dotted and the dash-dotted curves in
Figure~\ref{fig:p18scale} results from the isotropy of the overlapping
sphere model. The
isotropy condition $G(r)=G({\mathbf{r}})$ guarantees knowledge of the
full function $G({\mathbf{r}})$ from a measurement along any ray. Knowledge
of the full function is a prerequisite for calculating the three-dimensional
variance from (\ref{homo.var}). Hence, the above precedure is not applicable
for general anisotropic media.

\begin{figure}
\epsfysize=350pt.
\epsffile{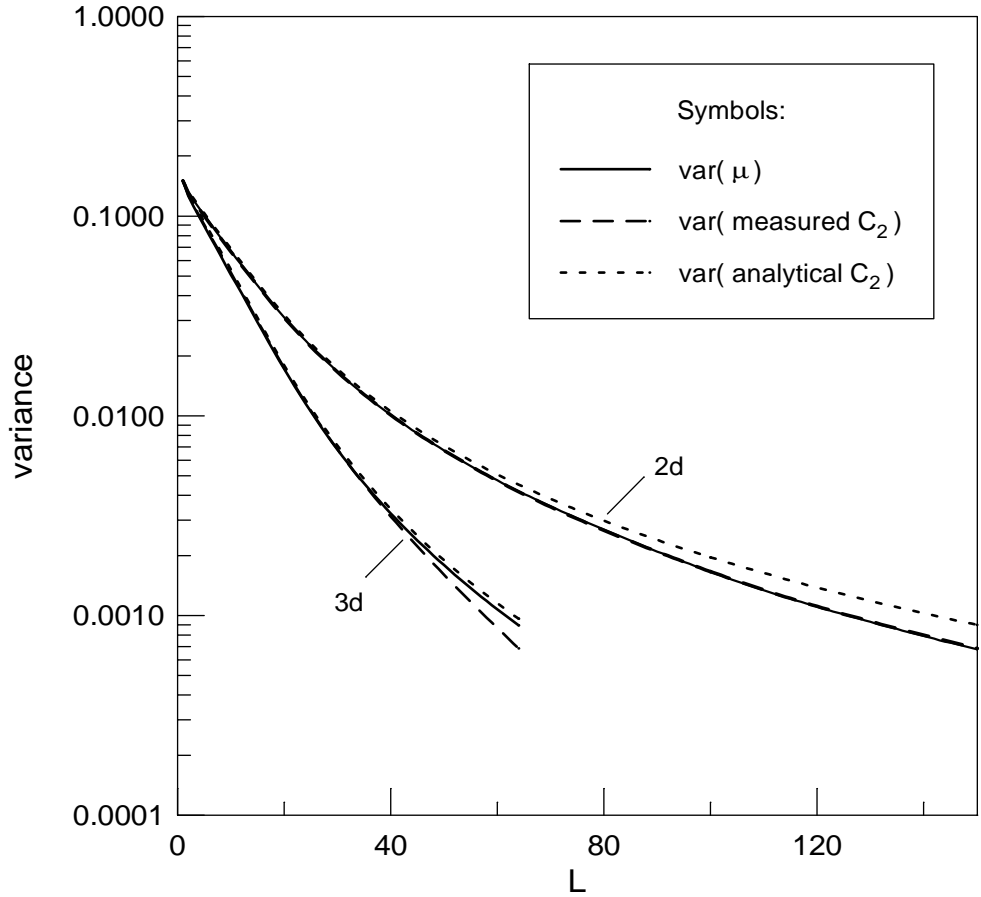}
\caption{Three different determinations of variance for both two- and 
three-dimensional LPDs.
The actual distribution-variances are the solid curves. The dashed and
dash-dotted curves are computed from (\protect\ref{homo.var}) using measured and
analytical values for $C_{2}$ repectively.}
\label{fig:p18var}
\end{figure}

To assess the relative reliability of our rescaling procedure for isotropic
systems, we compare in Figure~\ref{fig:p18var} the two- and three-dimensional
variances
computed from the exact and measured correlation functions $G(r)$ with
those computed from the local porosity distribution $\mu(\phi;{\mathbf{K}})$.
Solid lines are variances computed from $\mu(\phi;{\mathbf{K}})$, dashed
lines are variances computed from the measured $G(r)$ and dotted lines are
variances computed using the analytical $G(r)$. The overall agreement is
satisfactory, given that the side length of the whole system is only
256 voxels.

\section{A Sandstone specimen}
\begin{figure}
\epsfysize=400pt.
\epsffile{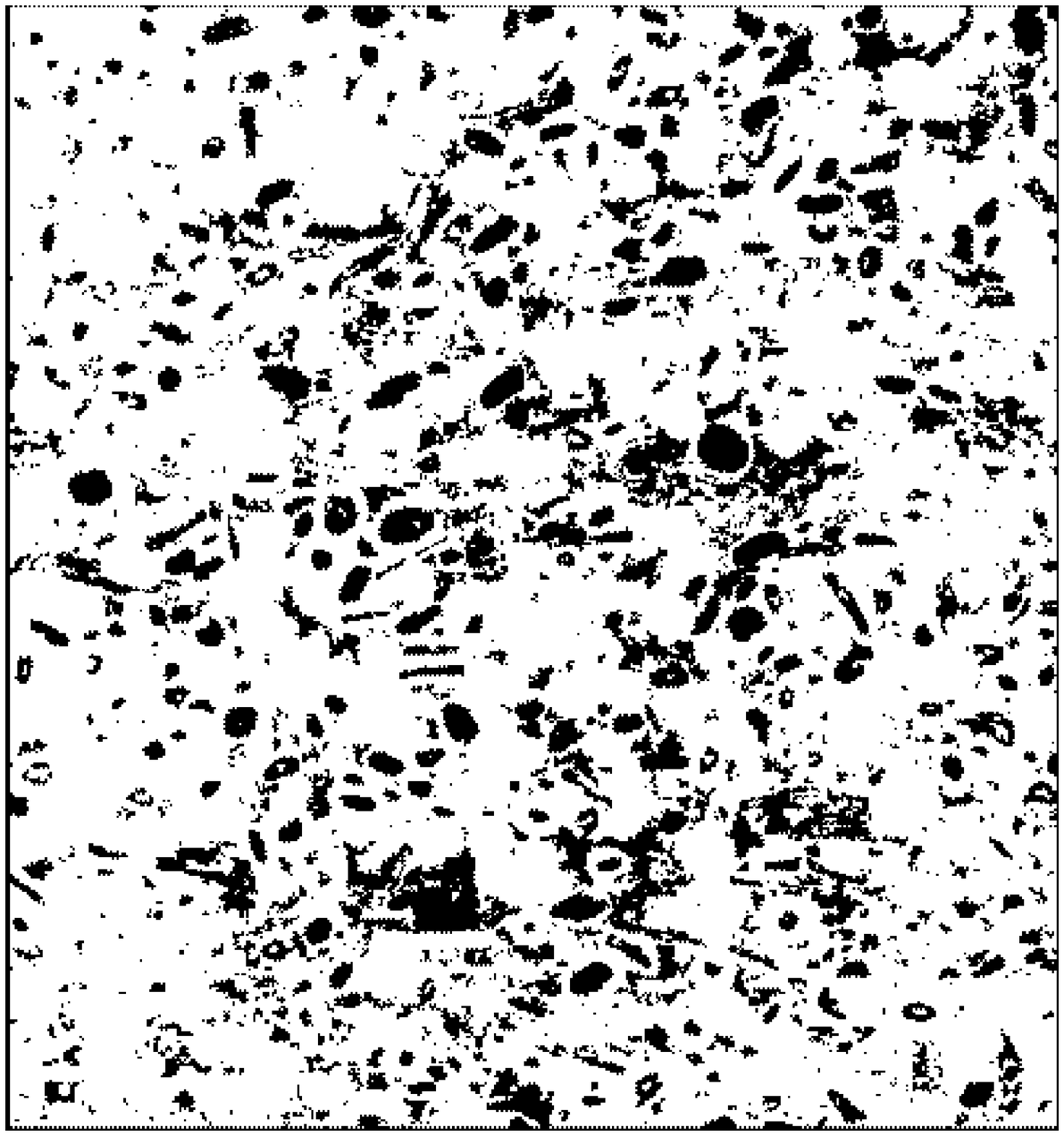}
\caption{Part of a typical planar-section through the Savonnier oolithic
sandstone. The side-length of the image is $\approx 1$ cm. The pore-space is
shown in black and rock matrix in white.}
\label{fig:Eslice}
\end{figure}

\begin{figure}
\epsfysize=320pt.
\epsffile{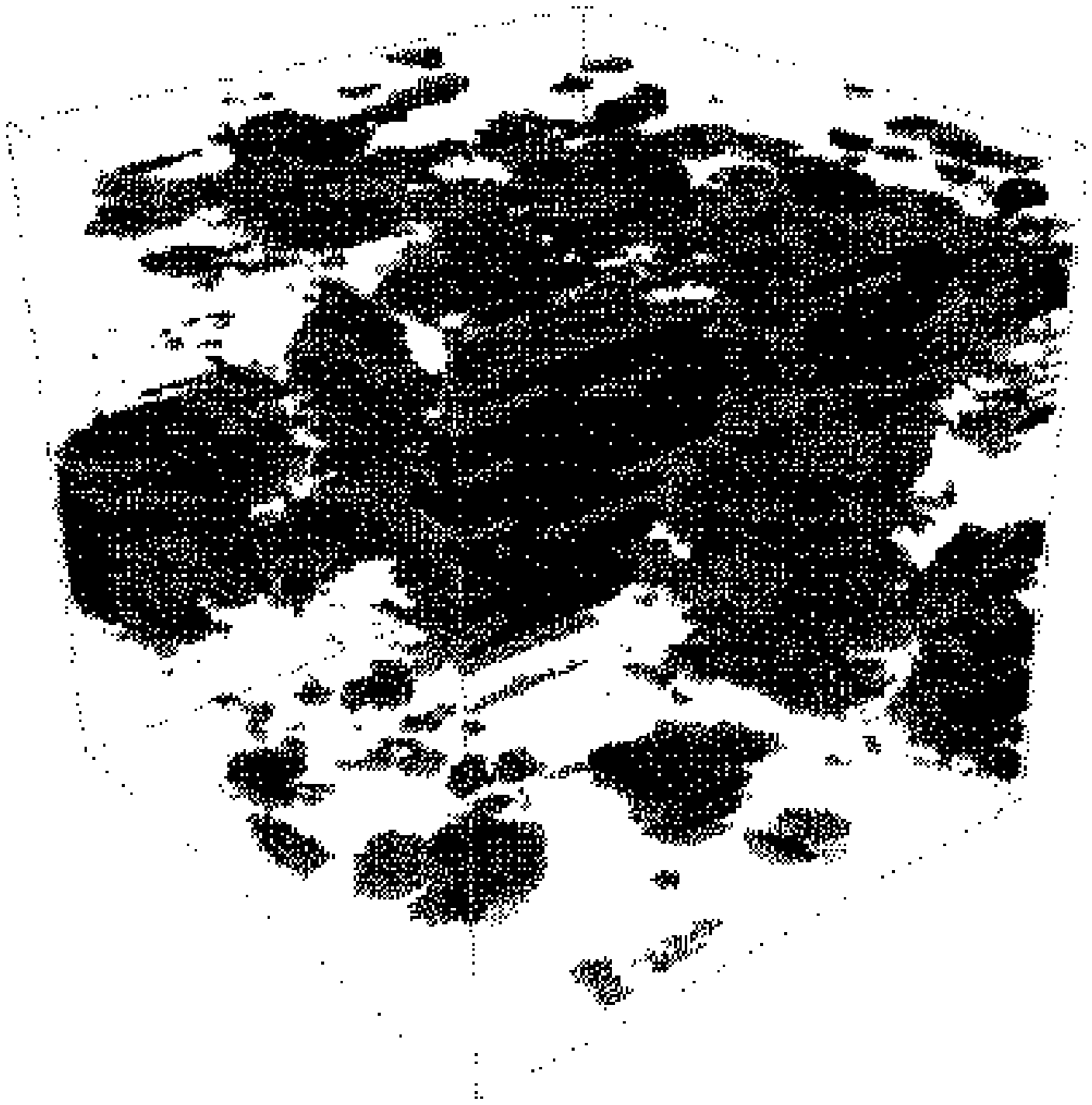}
\caption{A roughly cubic sub-volume of 100x100x99 voxels from the same
sandstone shown in the previous figure. The pore-space is shown in black,
matrix is transparent and the pore-space boundary is gray.}
\label{fig:Eblock}
\end{figure}

In this section, we analyse experimental data. The three-dimensional pore-space 
for a specimen of Savonnier oolithic sandstone was reconstructed from serial 
thin sections\cite{Ostertag}. 
The data-set consists of 99 binary images, each of 1904x1904 pixels, representing
planar cross-sections through the medium. The distance between the planes of 
consecutive images is $10\mu$m. This is also the pixel size. Each section was
obtained by thresholding a digitized micrograph into a black and white image.
The threshold value was chosen to match the experimentally-known bulk porosity
of $\overline{\phi} = 0.1845$. Figure~\ref{fig:Eslice} shows part of a typical 
image measuring approximately 1 cm along each side. In Figure~\ref{fig:Eblock} 
we show a three-dimensional
cubic subset with a side length of 1 mm, in which the pore-space is shown in
black, the matrix is transparent and the pore-space boundary is gray.

\begin{figure}
\epsfysize=350pt.
\epsffile{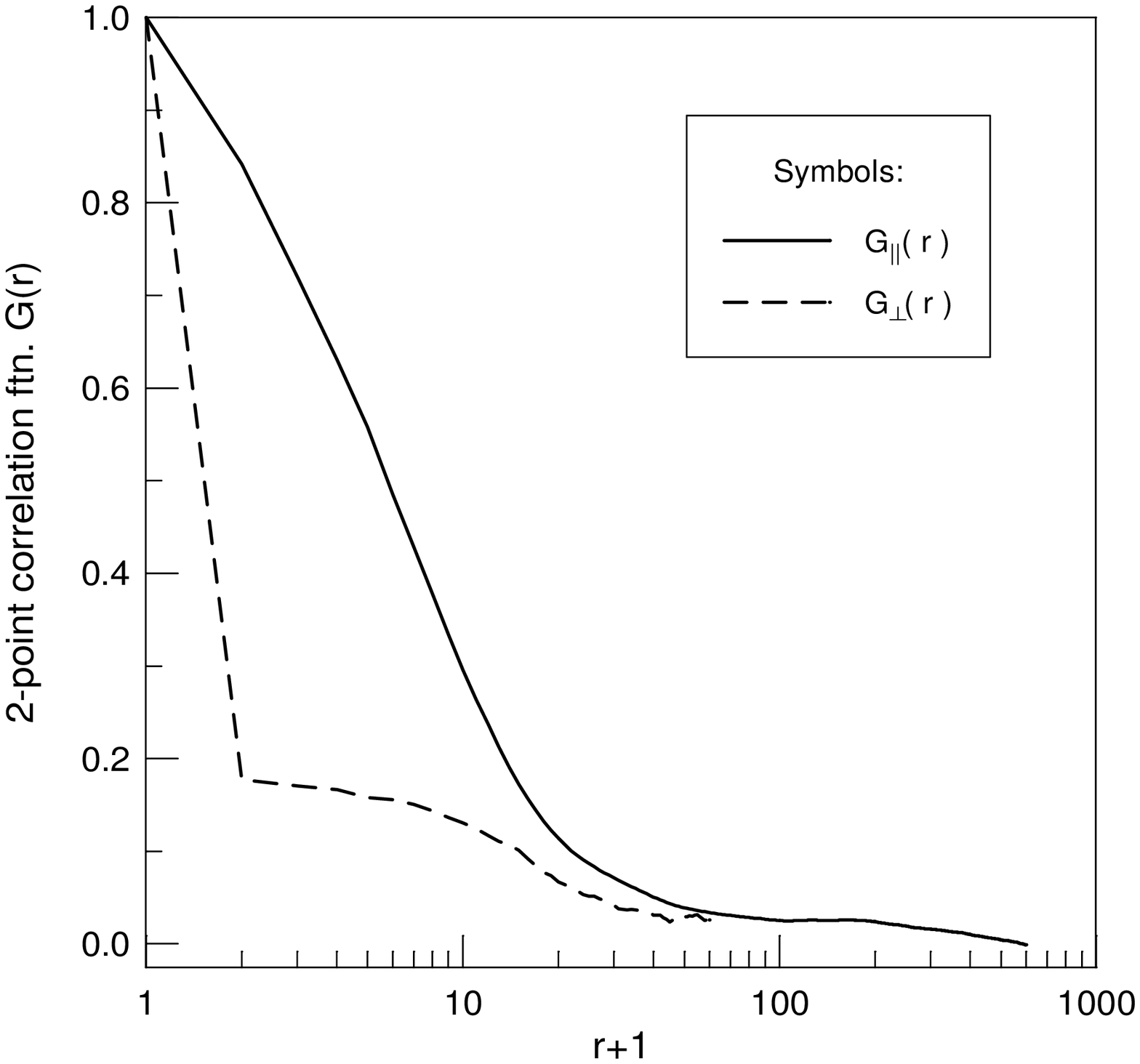}
\caption{Two-point correlation functions for the sandstone specimen. 
$G_{\parallel}(r)$
and $G_{\perp}(r)$ are shown as the solid and dashed curves respectively. These
functions are shown as functions of $r+1$ in order to plot the $r$-axis in
logarithmic scale. Note the presence of several correlation-lengths in 
$G_{\parallel}$.}
\label{fig:Eg}
\end{figure}

Figure~\ref{fig:Eblock} shows that the pore-space is anisotropic. This is 
corroborated in Figure~\ref{fig:Eg}, where the correlation function 
$G_{\parallel}(r)$, measured in the plane 
of the thin sections is shown together with $G_{\perp}(r)$ measured in the direction 
normal to the thin sections. The anisotropy is at least partly caused by the 
misalignments of individual images and variations in lighting and thresholding 
during the reconstruction. The correlation function $G_{\parallel}(r)$ also
shows the existence of large-scale heterogeneities. Given these imperfections
we do not expect to be able to reconstruct the three-dimensional LPDs from
two-dimensional information alone.

\begin{figure}
\epsfysize=460pt.
\epsffile{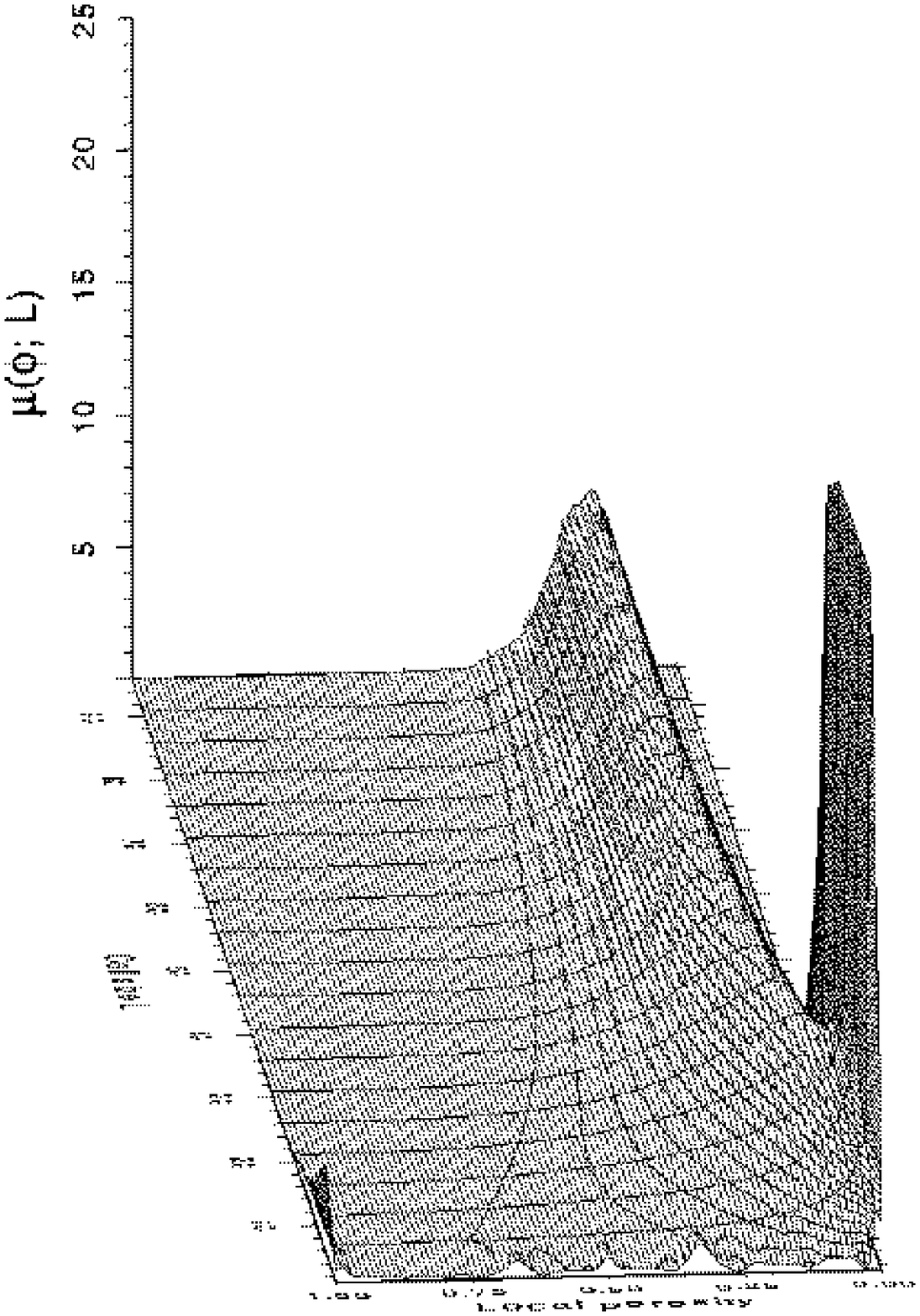}
\caption{The family of three-dimensional LPDs
$\mu_{3d}(\phi;L)$ for $L \in \{1,2, \ldots, 96\}$ for Savonnier 
oolithic sandstone. The spikes at regular intervals for small $L$ are 
not fluctuations, but rather a result of the anisotropy visible in
Figure~\ref{fig:Eblock} and Figure~\ref{fig:Eg}.}
\label{fig:E3dmu}
\end{figure}

\begin{figure}
\epsfysize=460pt.
\epsffile{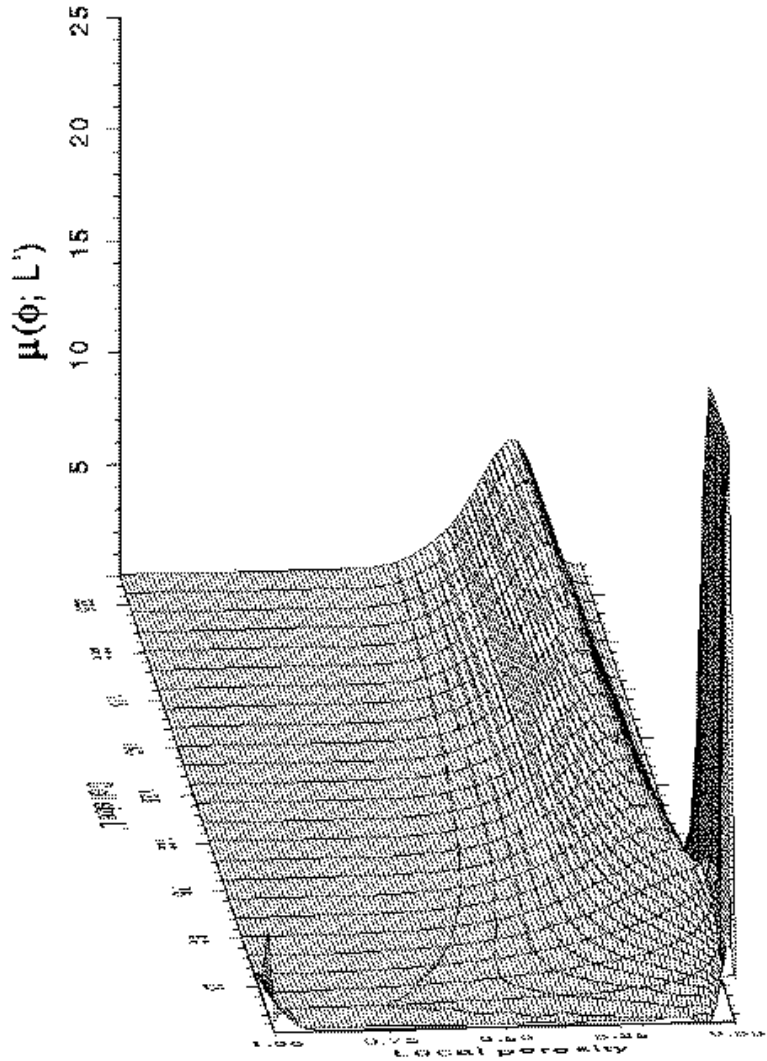}
\caption{The family of two-dimensional LPDs $\mu_{2d}(\phi;L~{'})$ for 
$L^{'} \in \{1,2, \ldots, 240\}$. Note that the distributions are much
broader than those for the overlapping sphere model shown in 
Figure~\ref{fig:p182dmu}.}
\label{fig:E2dmu}
\end{figure}

The family of three-dimensional LPDs is displayed in Figure~\ref{fig:E3dmu}.
The family of two-dimensional LPDs is displayed in Figure~\ref{fig:E2dmu}.
The three-dimensional distributions have been measured on a sub-system of
1881x1881x99 voxels for measurement cell sizes $L \in \{1,2,\ldots,96\}$.
The bulk porosity of the sub-system is 0.1861. The results displayed in
Figure~\ref{fig:E3dmu} are, to our knowledge, the first measurements of 
three-dimensional LPDs on a natural rock sample. The two-dimensional
LPDs shown for $L^{'} \in \{1,2,\ldots,240\}$ were obtained, as before, by
averaging the distributions for each plane.

\begin{figure}
\epsfysize=350pt.
\epsffile{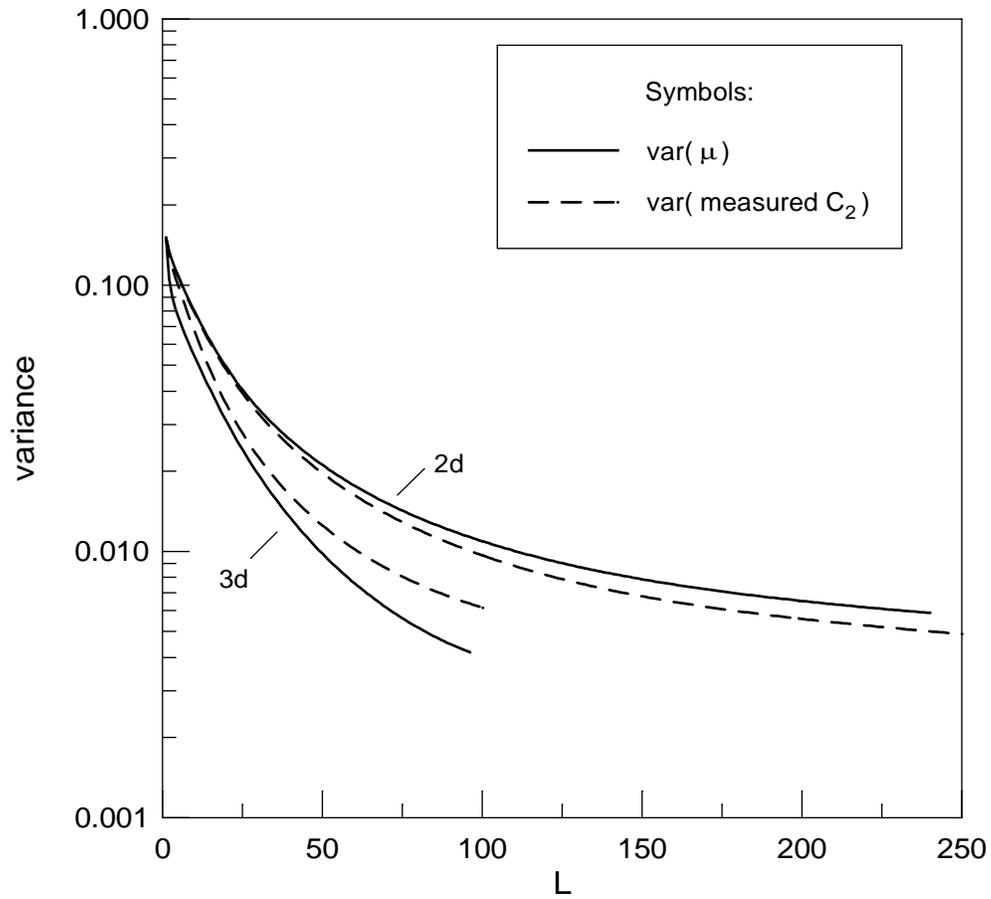}
\caption{Two- and three-dimensional local porosity variances for the
Savonnier oolithic sandstone. The solid curves are variances determined
directly from the distributions and the dashed curves are variances
computed from (\protect\ref{homo.var}) using the $G_{\parallel}(r)$.}
\label{fig:Evar}
\end{figure}

Figure~\ref{fig:Evar} shows the two- and three-dimensional variances computed 
directly
from the distributions as solid lines. The dashed lines are variances obtained
using $G_{\parallel}$ in (\ref{homo.var}). The poor agreement for $d=3$
is to be expected, since $G_{\parallel} \neq G_{\perp}$, as seen in 
Figure~\ref{fig:Eg}.
Hence, we cannot make an independent determination of the three-dimensional
variance from the two-dimensional data, and are unable to construct the
mapping $L^{'}(L)$ based on two-dimensional correlation functions.

\begin{figure}
\epsfysize=450pt.
\epsffile{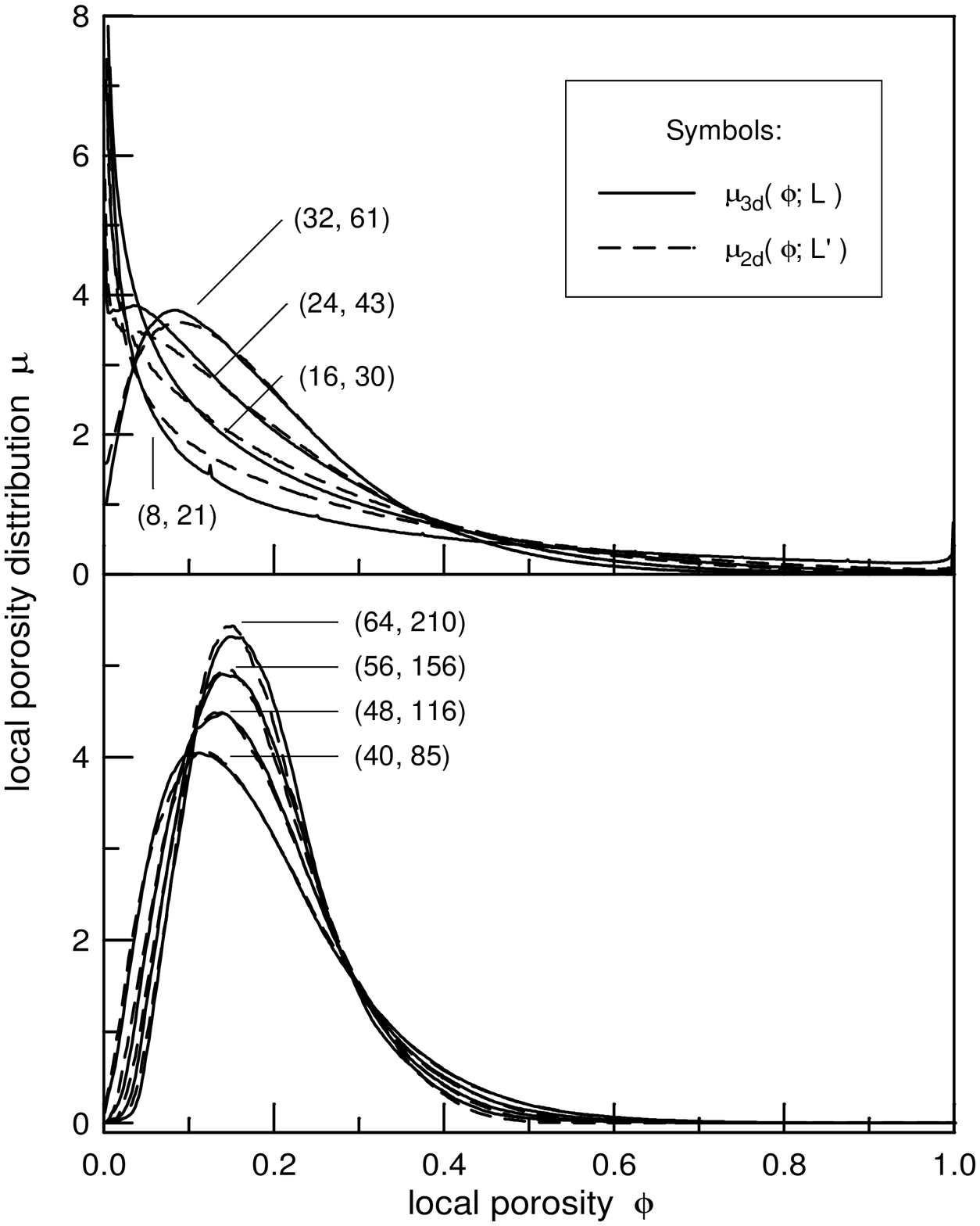}
\caption{$\Delta(L,L^{'})$-matched two- and three-dimensional LPDs for
the Savonnier
oolithic sandstone for a selection of $(L,L^{'})$-pairs indicated in the
figure. The two- and three-dimensional distributions are shown as dashed and
solid curves, respectively. The spikes in Figure~12a are seen here in
$\mu_{3d}(\phi;8)$ at $\phi = 1/8, 2/8$ and $3/8$.}
\label{fig:Ecorr}
\end{figure}

Nevertheless, a matching between two- and three-dimensional LPDs can be obtained
based on $\Delta(L,L^{'})$ as discussed above. The results of the 
$\Delta$-matching procedure are shown in Figure~\ref{fig:Ecorr} for a selection of 
$(L,L^{'})$-pairs indicated in the figure. The agreement is of similar quality 
as that for the overlapping sphere model. 

\section{Conclusion}
The present paper has investigated the dimensional dependence of local
porosity distributions, which formalize the idea of length-scale dependent
spatial averaging (coarse-graining). We have presented the first determinations
of LPDs for real porous media. We have shown that in general, in the limit of
large measurement cells(averaging regions), the two- and three-dimensional 
LPDs can be mapped onto each other by a rescaling of lengths $L^{'}(L)$.
This result was based on the central limit theorem and the concept of
stationarity. We have found that the mapping $L^{'}(L)$ extends below the
asymptotic regime to much smaller length-scales. This result is important for
both practical applications and because it agrees with the possibility of 
intermediate limit laws for LPDs based on generalizations of the central limit
theorem\cite{Hilfer4}.

An equally important result of this paper is the finding that the mapping 
$L^{'}(L)$ can be determined from purely two-dimensional measurements. However,
this result is restricted to homogeneous isotropic media.

We illustrated our results with three examples. For the percolation model,
the analytical form of the mapping $L^{'}(L)$ between two- and
three-dimensional LPDs is exactly known, and the agreement between mapped 
distributions becomes perfect. For the homogeneous isotropic overlapping 
sphere model, the agreement between mapped distributions is satisfactory for 
measurement cells of side $L \ge \xi$ and improves with increasing $L$. The 
mapping $L^{'}(L)$ can be successfully determined from two-dimensional 
observations alone for this medium. The three-dimensional pore-space 
reconstruction for the Savonnier oolithic sandstone specimen is clearly 
anisotropic and shows large-scale heterogeneities. The anisotropy of the
reconstructed pore-space does not allow one to construct the mapping
$L^{'}(L)$ from two-dimensional observations. Nevertheless, the mapped
curves from two- and three-dimensional measurements again show good
agreement.

\section{Acknowledgements}
We wish to thank C. Ostertag-Henning and Prof.\ Dr.\ R. Koch from the Dept.\ of 
Palaeontology at the Univ.\ of Erlangen for generously providing us with a 
three-dimensional pore-space reconstruction of a Savonnier oolithic sandstone
specimen. We also wish to thank Prof.\ Bjarne N{\o}st of the Dept.\ of Physics
at the Univ.\ of Oslo, and Dr.\ U. Mann and Prof.\ Dr.\ D. H. Welte at the
Forschungszentrum J{\"u}lich
for discussion. Two of us(R.H and E.H) wish to thank the Norges
Forskningsr{\aa}d for partial financial support.


\newpage

\begin{thebibliography}{9}                
\bibitem{Landauer}R Landauer, ``Electrical conductivity in inhomogeneous
media'', in {\it Electrical Transport and Optical Properties of Inhomogeneous
materials}(J. Garland and D. Tanner, eds.), (New York), p. 2, American Institute 
of Physics, 1978.
\bibitem{Mochan}W. Mochan and R. Barrera(eds.), {\it ETOPIM 3, Proceedings
of the Third International Conference on Electrical Transport and Optical
Properties of Inhomogeneous Media}, vol. Physica A 207. Amsterdam: North
Holland, 1994.
\bibitem{Hilfer4}R. Hilfer, ``Transport and Relaxation in Porous Media'',
{\it Adv. Chem. Phys.}, vol. XCII, p. 299, 1996, in print.
\bibitem{Sahimi}M. Sahimi, ``Flow phemomena in rocks: From continuum
models fractals, percolation, celullar automata and simulated annealing'',
{\it Rev. Mod. Phys.}, vol. 65, p. 1393, 1993.
\bibitem{Lafait}J. Lafait and D. Tanner(eds.), {\it ETOPIM 2, Proceedings 
of the Second International Conference on Electrical and Optical Properties
of Inhomogeneous Media}, vol. Physica A 157. Amsterdam: North Holland, 1989.
\bibitem{Debye1}P.Debye and A. Bueche, ``Scattering by an inhomogeneous
solid'', {\it L. Appl. Phys.}, vol. 20, p. 518, 1949
\bibitem{Debye2}P.Debye and A. Bueche, ``Scattering by an inhomogeneous
solid II: The correlation function and its application'', {\it L. Appl. Phys.},
vol. 28, p. 679, 1957
\bibitem{Weissberg}H. Weissberg, ``Effective diffusion coefficient in
porous media'', {\it J. Appl. Phys.}, vol. 34, p. 2636, 1963.
\bibitem{Torquato1}S. Torquato and G. Stell, ``Microstructure of Two
Phase Random Media I: The $n$-Point Probability Functions'', {\it J.
Chem. Phys.}, vol. 77, p. 2071, 1982.
\bibitem{Torquato2}S. Torquato and G. Stell, ``Microstructure of Two
Phase Random Media II: The Mayer-Montroll and Kirkwood-Salsburg
Heirarchies'', {\it J. Chem. Phys.}, vol. 78, p. 3262, 1983.
\bibitem{Rikvold1}P. Rikvold and G. Stell, ``Porosity and specific
surface for impenetrable-sphere models of two-phase random media'', 
{\it J. Chem. Phys.}, vol. 82, p. 1014, 1985.
\bibitem{Rikvold2}P. Rikvold and G. Stell, ``D-dimensional
interpenetrable sphere models of random two-phase media:
Microstructure and applications to chromotography'', {\it J. Colloid
and Interface Sci.}, vol. 108, p. 158, 1985.
\bibitem{Stell}G. Stell and P. Rikvold, ``Polydispersity in fluids and
composites: Some theoretical results'', {\it Int. J. Thermophysics},
vol. 7, p. 863, 1986.
\bibitem{Berryman1}J. Berryman, ``Measurement of spatial correlation
functions using image processing techniques'', {\it J. Appl. Phys.},
vol. 57, p. 2374, 1985
\bibitem{Berryman2}J. Berryman and S. Blair, ``Use of digital image
analysis to estimate fluid permeability of porous media: Application
of two-point correlation functions'', {\it J. Appl. Phys.},
vol. 60, p. 1930, 1986
\bibitem{Hilfer1}R. Hilfer, ``Geometric and dielectric 
charaterizations of porous media'', {\it Phys. Rev. B}, vol. 44, p. 60,                
1991.                
\bibitem{Boger}F.Boger, J. Feder, R. Hilfer, and T. J{\o}ssang, 
``Microstructural sensitivity of local porosity distributions'',
{\it Physica A}, vol. 187, p. 55, 1992.
\bibitem{Hilfer2}R. Hilfer, ``Local porosity theory for flow in 
porous media'', {\it Phys. Rev. B}, vol. 45, p. 7115, 1992.                
\bibitem{Hilfer3}R. Hilfer, B. N{\o}st, E. Haslund, T. Kautzsch, B. Virgin
and B. D. Hansen, ``Local porosity theory for the frequency dependent
dielectric function of porous rocks and polymer blends'',
{\it Physica A}, vol. 207, p. 19, 1994.
\bibitem{Haslund}E.Haslund, B. D. Hansen, R. Hilfer, and B. N{\o}st,
``Measurements of local porosities and dielectric dispersion for a 
water saturated porous medium'', {\it J. Appl. Phys.}, vol. 76,
p. 5473, 1994. 
\bibitem{O'Neill}E. O'Neill, {\it Introduction to Statistical Optics},
Reading: Addisons-Wesley, 1963. 
\bibitem{Coker} D. A. Coker, S. Torquato, ``Extraction of
morphological quantities from a digitized medium'', {\it J. Appl.
Phys.}, vol. 77, p. 6087, 1995.
\bibitem{Ostertag}C. Ostertag-Henning, B. Virgin, T. Rage, R. Hilfer,
R. Koch and U. Mann, ``Messung dreidimensionaler lokaler
Porositi\"{a}tsverteilungen'', 1995, unpublished.
\end{thebibliography}
\end{document}